\documentclass[sigconf]{acmart}
\AtBeginDocument{%
  }

\copyrightyear{2026}
\acmYear{2026}
\setcopyright{cc}
\setcctype{by}
\acmConference[CHI EA '26]{Extended Abstracts of the 2026 CHI Conference on Human Factors in Computing Systems}{April 13--17, 2026}{Barcelona, Spain}
\acmBooktitle{Extended Abstracts of the 2026 CHI Conference on Human Factors in Computing Systems (CHI EA '26), April 13--17, 2026, Barcelona, Spain}
\acmDOI{10.1145/3772363.3798977}
\acmISBN{979-8-4007-2281-3/2026/04}

\usepackage{subcaption}
\usepackage{enumitem}

\begin{document}

\title{Exploring Experiential Differences Between Virtual and Physical Memory-Linked Objects in Extended Reality}

\author{Zaid Ahmed}
\orcid{0009-0007-6307-9229}
\affiliation{%
  \institution{University of Calgary}
  \city{Calgary}
  \state{Alberta}
  \country{Canada}
}
\email{zaid.ahmed1@ucalgary.ca}

\author{Omar A.\ Khan}
\orcid{0009-0000-1705-0950}
\affiliation{%
  \institution{Drexel University}
  \city{Philadelphia}
  \state{Pennsylvania}
  \country{United States}
}
\email{omar.ahmad.khan@drexel.edu}

\author{Hyeongil Nam}
\orcid{0000-0002-6017-8869}
\affiliation{%
  \institution{University of Calgary}
  \city{Calgary}
  \state{Alberta}
  \country{Canada}
}
\email{hyeongil.nam@ucalgary.ca}

\author{Kangsoo Kim}
\authornote{Corresponding author.}
\orcid{0000-0002-0925-378X}
\affiliation{%
  \institution{University of Calgary}
  \city{Calgary}
  \state{Alberta}
  \country{Canada}
}
\email{kangsoo.kim@ucalgary.ca}

\renewcommand{\shortauthors}{Ahmed et al.}

\begin{abstract}
  Extended Reality (XR) enables immersive capture and re-experience of personal memories, yet how interface representations shape these experiences remains underexplored.
  We examine how users relive and share XR memories through three interaction approaches: (1) physical memory-linked objects, (2) virtual memory-linked objects, and (3) a conventional virtual gallery interface.
  In a within-subjects study ($N=24$, 12 pairs), participants captured shared experiences using 360° video and later accessed and shared these memories across the three interfaces.
  We analyzed open-ended qualitative responses focusing on perceived value, enjoyment, usability, emotional attachment, and social connection.
  The findings reveal trade-offs: physical objects fostered stronger social connection and conversation through tangible exchange; virtual objects balanced engagement and usability; and the gallery interface was efficient but less personal.
  These results suggest that object-based representations---physical and virtual---support key social dimensions of XR memory experiences, offering lessons for designing future systems that emphasize shared meaning and interpersonal connection.
\end{abstract}


\begin{teaserfigure}
  \includegraphics[width=\textwidth]{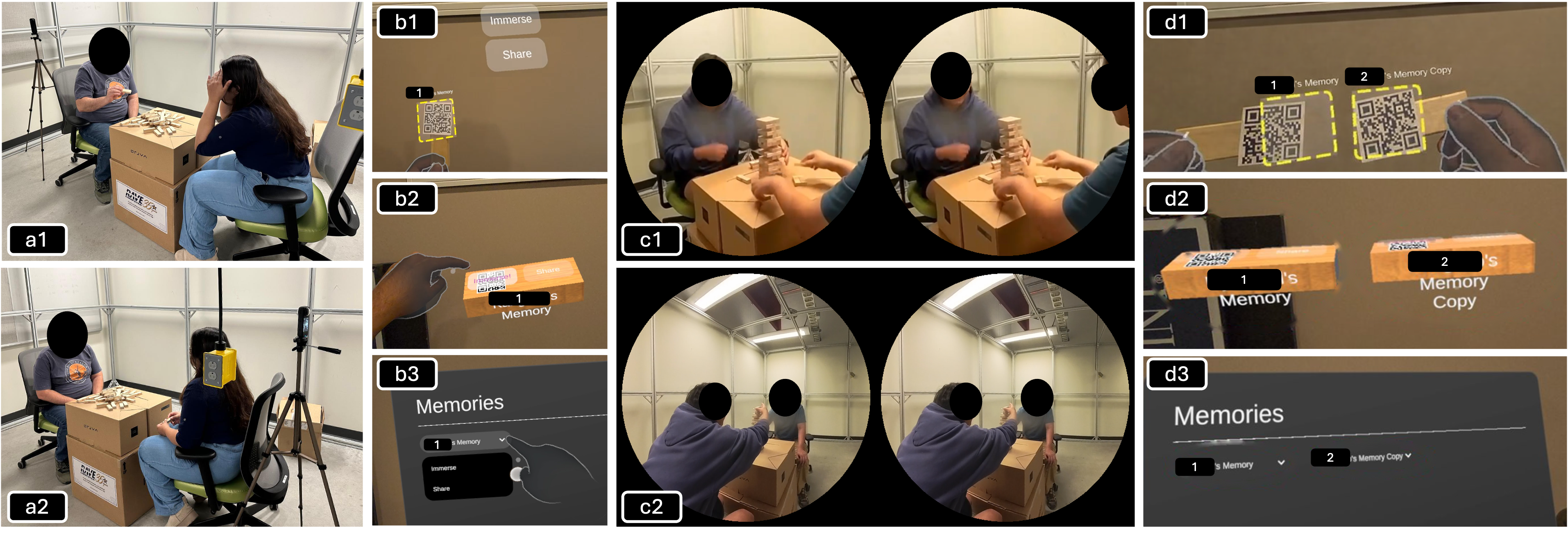}
  \vspace{-6ex}
  \caption{Overview of the study workflow and three memory interaction interfaces. (a1--a2) Capture: a 360° camera records participants playing Jenga; (b1--b3) Interface: physical block, virtual block, and virtual gallery interfaces; (c1--c2) Relive: replaying the stored memory in an immersive virtual environment; and (d1--d3) Share: transferring and accessing memories across the three interfaces.}
  \Description{A composite figure showing the study workflow across four stages. Stage (a) shows two participants sitting at a table playing Jenga with a 360-degree camera behind them capturing the experience. Stage (b) shows three interface conditions side by side: b1 shows a physical Jenga block with a QR code sticker being held in front of a VR headset displaying Immerse and Share buttons; b2 shows a virtual Jenga block floating in the VR environment with Immerse and Share buttons on its surface; b3 shows a virtual gallery window with a dropdown menu containing memory options. Stage (c) shows the immersive 360-degree video playback of the Jenga game as seen from inside the VR headset. Stage (d) shows memory sharing for each interface: d1 shows a physical block being handed between participants; d2 shows a virtual block appearing on the receiving participant's screen; d3 shows the shared memory appearing in the gallery dropdown list.}
  \label{fig:teaser}
\end{teaserfigure}

\maketitle

\section{Introduction}

Extended Reality (XR) offers new possibilities for storing and reliving personal and social memories in immersive ways~\cite{xi_virtual_2024}~\cite{kim_memento_2026}.
Unlike photographs or conventional video, XR media such as 360° video and spatial reconstructions allow users to revisit past moments with a heightened sense of presence, enabling them to look around, re-experience spatial context, and feel situated within the original event~\cite{ventura_immersive_2019}.
As XR memory capture becomes more accessible, an important challenge shifts from how memories are recorded to how interaction techniques and interface designs shape the ways memories are accessed, experienced, and shared.

As XR memory systems mature, a key design question concerns how users should access and interact with these memories through \emph{memory-linked objects}---object-based interfaces that function as interactive memory cues.
Prior work has explored a range of interaction paradigms, including conventional gallery-style interfaces as well as tangible user interfaces (TUIs) that link digital content to physical objects~\cite{ishii_tangible_1997,mugellini_using_2007}.
Each approach introduces distinct benefits and trade-offs.
Physical tangible interfaces can support embodied interaction, engagement, and collaboration by leveraging users' familiarity with real-world objects~\cite{shaer_tangible_2009}. 
The interweaving of material and social aspects in tangible interaction~\cite{hornecker_getting_2006} makes such interfaces particularly relevant for shared memory experiences. 
Physical artifacts have been shown to act as powerful memory triggers, helping bridge the gap between physical and digital memories~\cite{van_den_hoven_informing_2008}. 
Research on physical vs digital mementos has found that physical objects are highly valued and support rich recollection, while digital mementos, though offering practical advantages, are often perceived as less valuable and infrequently accessed~\cite{petrelli_family_2010}.

However, TUIs also present practical challenges, including the need to maintain dedicated physical artifacts and limited scalability as the number of digital items grows~\cite{shaer_tangible_2009}.
Virtual object-based interfaces may offer similar affordances to physical tangibles in certain contexts without requiring material infrastructure~\cite{cuendet_tangible_2012}. How these different interface representations shape users' experiences with XR memories remains underexplored.

Rather than assuming inherent advantages of any single approach, this paper adopts an exploratory, comparative perspective on XR social memory interaction extending the concept of \emph{``TangibleMoments''} proposed by Khan et al.~\cite{Khan2025tangible}. 
We investigate how different interface representations---(1) physical memory-linked objects, (2) virtual memory-linked objects, and (3) a conventional virtual gallery interface---shape users' experiences when capturing, reliving, and sharing immersive memories.
To this end, we conduct a within-subjects user study ($N=24$, 12 pairs) comparing these three interface conditions in a shared memory context, focusing on the experiential outcomes including perceived value, enjoyment, usability, emotional
attachment to the memory objects, and social bonding between participants.

Through this comparative study, we aim to inform the design of future XR memory systems by clarifying how physical object-based, virtual object-based, and gallery-based interfaces support different aspects of memory experience, including qualitative feedback on perceived value, enjoyment, usability, emotional attachment, and social bonding. 
By examining both virtual and physical objects alongside a more traditional interface, this work contributes empirical insight into how representation choices shape XR memory interaction beyond assumed benefits of tangibility alone.

\section{Materials}

\subsection{Memory Experience with Virtual and Physical Memory-Linked Objects}

In our study, participants engaged in two rounds of Jenga gameplay while a 360-degree camera was positioned behind them to capture immersive memories of the experience (Figure~\ref{fig:teaser}(a)).
These captured memories were subsequently loaded onto an HMD, enabling participants to relive and share their experiences through three different interaction modalities, presented in counterbalanced order.

\paragraph{\textbf{Physical Block}}

The tangible physical block interface combined physical and virtual elements to facilitate memory retrieval.
Participants affixed a QR code sticker to a physical Jenga block and held it within their field of view (FOV).
The system automatically detected and scanned the QR code, triggering the appearance of two interactive buttons: \textit{Immerse} and \textit{Share} (Figure~\ref{fig:teaser}(b1)).
To interact with the system, participants used direct touch input by physically poking at the desired button. 
Selecting \textit{Immerse} initiated playback of the corresponding 360-degree video, immersing the participant in their recorded memory (Figure~\ref{fig:teaser}(c).
Once playback concluded, two options appeared: \textit{Replay} to view the memory again, or \textit{Go Back} to return to the main interface.
The \textit{Share} functionality required participants to hold up a second physical block with a new QR code, initiating the transfer process from the source block to the destination block.
After scanning the second block, participants pressed a \textit{Confirm} button to complete the transfer.
The physical block was then handed to the other participant, who pressed a \textit{Sync} button on their screen to refresh the database. 
Once synchronized, they could scan the received block to relive the shared memory (Figure~\ref{fig:teaser}(d1)).

\paragraph{\textbf{Virtual Block}}

The virtual block interface displayed a virtual Jenga block containing the memory (Figure~\ref{fig:teaser}(b2)).
Participants could manipulate the block's position by pinching a corner, moving it to the desired location, and releasing to fix its position.
The block featured two buttons directly on its surface: \textit{Immerse} and \textit{Share}.
The immersion experience mirrored that of the physical block, playing the 360-degree video upon selection (Figure 1(c)).
The \textit{Share} functionality presented a confirmation dialog. 
After pressing \textit{Confirm Share}, the other participant could press \textit{Sync}, causing a virtual block containing the shared memory to appear on their screen (Figure~\ref{fig:teaser}(d2)).

\paragraph{\textbf{Virtual Gallery}}

The gallery interface presented participants with a virtual gallery window containing their memories as dropdown options (Figure~\ref{fig:teaser}(b3)).
Participants could interact with the interface using two input methods: (1) direct finger poking to press buttons, or (2) a pinch gesture where the thumb and forefinger were held several centimeters apart to position a cursor, then pinched together to click.
Upon clicking the dropdown menu, two options appeared: \textit{Immerse} and \textit{Share}. 
Selecting \textit{Immerse} played the 360-degree video of their captured memory, concluding with \textit{Replay} and \textit{Go Back} options (Figure~\ref{fig:teaser}(c)).
The \textit{Share} functionality presented the same confirmation dialog as the virtual block interface.
After pressing \textit{Confirm Share}, the other participant could press \textit{Sync} on their screen to refresh the interface and access the shared memory within their gallery (Figure~\ref{fig:teaser}(d3)).

\paragraph{\textbf{Common Immersion Experience}}

Across all three systems, the \textit{Immerse} function provided a consistent experience. 
Upon selection, the system automatically initiated playback of the 360-degree video captured during the participant's Jenga gameplay. 
The video played in full, after which the interface presented two options: \textit{Replay} to review the memory again, or \textit{Go Back} to return to the system's main interface.

\subsection{Prototype Implementation}
\label{Sec:Prototype}

To demonstrate the potential of memory-linked objects, we developed an initial prototype.
The prototype was built using the Unity game engine (v6000.0.39f1) with the Meta XR All-in-One SDK and developed for the Meta Quest 3 HMD. 
The stickers displayed QR codes representing unique identifiers linked to particular memories stored in our system.
For QR code detection and scanning functionality, we leveraged the passthrough camera API samples provided by Meta\footnote{\url{https://github.com/oculus-samples/Unity-PassthroughCameraApiSamples} (Accessed 2025-05-01)}.

The XR memories in our prototype consisted of 360-degree videos captured during participant gameplay sessions. 
We used an Insta360 X3 camera positioned behind participants to record immersive footage of their Jenga gameplay. 
The recorded videos were subsequently processed using the Insta360 Studio application to create short video clips suitable for playback within the VR environment.

\section{Methods}
\label{Sec:Experiment}

\subsection{Participants}
Twenty-four participants (14 male, 10 female)
were recruited in 12 pairs through social media, word of mouth, and physical posters.
All pairs had pre-existing relationships: 10 pairs were friends, one pair were siblings, and one pair showed mixed responses (one participant identified the other as a friend, while the other identified them as an acquaintance).
Prior VR experience varied, with most participants (15) having used VR only a few times, six having no prior experience, and three reporting regular use (weekly or monthly). 
Relationship closeness was assessed using the Inclusion of Other in the Self (IOS) scale~\cite{aron_inclusion_1992}, a single-item pictorial measure where respondents select one of seven pairs of increasingly overlapping circles in response to the prompt ``Circle the picture which best describes your relationship with the other participant'' (1 = no overlap, 7 = almost concentric). 
Scores ranged from 1 to 6 with a mean of 3.25.
Each participant received an Amazon gift card (\$15 CAD) as compensation.
This study was approved by the institutional ethics review board of the University of Calgary Conjoint Faculties Research Ethics Board (\#REB25-0232).

\subsection{Study Design}

This study used a within-subjects design to compare three interface conditions for accessing and sharing immersive 360° video memories: (1) a physical block interface, where participants scanned QR codes embedded attached to Jenga blocks; (2) a virtual block interface, where participants interacted with virtual representations of Jenga blocks in XR to access the same content; and (3) a gallery interface, where participants selected memories through a traditional 2D UI panel displayed within the headset.
With three conditions, there were six possible orderings; we conducted 12 sessions (two per order) to fully counterbalance condition presentation.

\subsection{Procedure}

Each session involved a pair of participants with a pre-existing relationship. 
After arrival and consent, participants first played two games of Jenga together while being recorded by a 360° camera.
In each game, the camera was positioned behind a different participant. 
From each recording, we extracted the final 10 seconds surrounding the tower's collapse to serve as the memory clip for that participant for the remainder of the study.
Before beginning the experimental conditions, participants completed a tutorial scene designed to familiarize them with VR interaction.
The tutorial included four activities: scanning QR codes, poking buttons, moving virtual cubes, and interacting with UI panels. 
These activities mirrored the interactions required across the three experimental conditions.
Participants then completed all three interface conditions. 
Before each condition, they viewed a brief slide deck demonstrating the actions for reliving and sharing memories with that interface. 
Following all three conditions, participants completed an open-ended questionnaire comparing their experiences across
interfaces.
The study duration was approximately 1 hour.

\subsection{Open-ended Questionnaire}

To analyze participants' qualitative feedback on perceived value, enjoyment, usability, emotional attachment, and social connection, we used an open-ended questionnaire consisting of ten questions listed below.
\begin{enumerate}[nosep]
\small
    \item Which of the three interfaces (physical block, virtual block, gallery) felt the most \emph{meaningful} and/or \emph{valuable} and why?
    \item Which of the three interfaces was the most \emph{enjoyable} and/or \emph{fun}, and why?
    \item Which of the three interfaces was the \emph{easiest} and/or most \emph{intuitive} to use, and why?
    \item Which of the three interfaces made you feel the most \emph{socially connected} with your co-participant, and why?
    \item Did any interface \emph{encourage more conversation} or shared storytelling compared to the others? If so, which one and why?
    \item Which of the three interfaces did you \emph{prefer for reliving} the memories, and why?
    \item Which did you \emph{prefer for sharing} the memories, and why?
    \item Which was overall the \emph{best}, and why?
    \item If you were to use a similar system in real life to revisit personal memories, which interface would you \emph{use} and why?
    \item Do you have any suggestions for improving any aspect of any of the interfaces?
\end{enumerate}
Using inductive thematic analysis, the first author coded the open-ended responses to identify recurring themes.
The results were reviewed with the co-authors to cross-check interpretations against the raw data and resolve discrepancies through discussion. In the cases where participants explicitly stated no preference or expressed equal preference for multiple interfaces, all named interfaces were counted.

\section{Results and Discussion}

Responses were grouped by selected interface for each question (Table~\ref{tab:qualitative_summary}); results reflect self-reported perceptions rather than objective performance measures.


\begin{table}[t]
\centering
\small
\caption{Distribution of interface preferences across qualitative questions (N=24).}
\vspace{-1em}
\label{tab:qualitative_summary}
\begin{tabular}{lccc}
\hline
\textbf{Question} & \textbf{Physical} & \textbf{Virtual} & \textbf{Gallery} \\
\hline
Most meaningful/valuable & 10 & \textbf{13} & 7 \\
Most enjoyable/fun & 8 & \textbf{18} & 5 \\
Easiest/most intuitive & 3 & 6 & \textbf{19} \\
Most socially connected & \textbf{20} & 9 & 5 \\
Encouraged conversation & \textbf{18} & 9 & 4 \\
Preferred for reliving & 6 & \textbf{17} & 11 \\
Preferred for sharing & 8 & 8 & \textbf{10} \\
Overall best & 5 & \textbf{14} & 8 \\
Real-life use & 6 & \textbf{12} & 8 \\
\hline
\textbf{Total} & 84 & \textbf{106} & 77 \\
\hline
\end{tabular}
\vspace{-2ex}
\end{table}

\subsection{Physical Block: Social Connection Through Tangible Exchange}
The physical block interface demonstrated a clear advantage for social connection and conversation.
When asked which interface made them feel most socially connected, 20 of 24 participants (83\%) selected physical blocks, and 18 (75\%) identified it as encouraging more conversation.
Participants consistently emphasized the significance of the tangible exchange, noting that the act of physically handing over a block created a shared experience that felt meaningful.
One participant explained: \textit{``I felt most socially connected by the physical block because of the actual exchange of a physical item. The other ones you don't actually interact with the co-participant so it felt like my own experience but the action of physically copying my memory and giving it to the other person felt like we were sharing something.''} (P20).
As another participant noted: \textit{``I felt most connected to the physical jenga block because the idea that the jenga block held the memory of us playing jenga felt meaningful and almost sentimental.''} (P17).

\subsection{Virtual Block: Balancing Engagement and Usability}
The virtual block interface emerged as the overall preferred option, selected by 14 participants (58\%) as the best interface overall. 
It was also the most preferred for reliving memories (17 participants, 71\%) and rated most enjoyable by 18 participants (75\%). 
Participants appreciated that virtual blocks combined the object-based interaction of physical blocks with improved usability.
As one participant explained: \textit{``The virtual block combined the strengths of the other two, with less of the drawbacks.''} (P12). 
Another noted: \textit{``Virtual block was the best due to [the] ease of use which we have in gallery and it also consisted of a block which we had in physical block... a nice mix of both.''} (P4).

The novelty and interactivity of manipulating virtual objects contributed to enjoyment.
Participants described the experience as \textit{``fascinating''} (P10) and compared it to science fiction interfaces: \textit{``I was feeling like I am doing what Iron Man does in Avengers movie.''} (P13).
Participants perceived the symbolic connection between the virtual blocks and the memory content as enhancing the experience: \textit{``The virtual block felt the most meaningful because it was easy to use and the block symbolized the activity that we did and reminded me of the good memory.''} (P18).

Importantly, virtual blocks were also perceived to offer some of the social benefits associated with physical blocks.
Nine participants (38\%) rated virtual blocks highest for social connection, and participants noted its potential for remote sharing: \textit{``The virtual block would be the best way for sharing memories as they reminded me of the game we played together and also could be useful if my friend was further away from me.''} (P17).

\subsection{Virtual Gallery: Familiar and Efficient}
The gallery was rated easiest and most intuitive by 19 participants (79\%), reflecting its alignment with familiar digital interfaces. 
Participants appreciated the straightforward interaction: \textit{``The gallery was the easiest and most intuitive to use, mainly due to having less interactions to go through in order to accomplish the same functionality. Everything is laid out in front of you.''} (P12).
The familiarity of the interface reduced cognitive load: 
\textit{``I have seen gallery interfaces on my phone and laptops already and I probably would have figured my way out without instructions.''} (P4).

For sharing memories, the gallery was slightly preferred (10 participants, 42\%) over the other interfaces, with participants noting its efficiency: \textit{``If I want to use it for everyday use, 
I would use gallery because it is faster and easier to use than the others.''} (P22).
However, participants also noted that this efficiency came at the cost of personal connection: \textit{``The gallery interface was the easiest to use because it was very similar to something that you see on a day-to-day basis like a computer interface.''} (P16), but another observed that \textit{``it felt the least personal.''} (P23).


\section{Conclusions: Findings and Implications}

We compared three interface representations for reliving and sharing immersive XR memories---physical objects, virtual objects, and a conventional gallery---using a qualitative, within-subjects study.
The results show clear experiential trade-offs: physical objects were reported to most strongly support social connection and conversation through tangible exchange;
virtual objects were perceived to balance engagement, usability, and symbolic meaning, leading to the highest overall preference; and the gallery was perceived as the most efficient and intuitive but least personal.

These findings suggest that the benefits often attributed to physical tangibility are not limited to material artifacts.
Well-designed virtual object representations may preserve key social and emotional qualities of physical interaction while avoiding practical limitations of physical systems.
For XR memory system design, object-based representations---physical or virtual---offer advantages over purely gallery-based interfaces when supporting shared meaning and interpersonal connection, particularly in social memory contexts.
Future work should examine whether these patterns generalize to other types of memories, such as travel, family events, or co-creative experiences, beyond the Jenga-based scenario.

\begin{acks}
We acknowledge the support of the Natural Sciences and Engineering Research Council of Canada (NSERC), [RGPIN-2022-03294].
\end{acks}

\bibliographystyle{ACM-Reference-Format}
\bibliography{template}

\end{document}